\documentclass[%
 reprint,
 amsmath,amssymb,
 aps,
 onecolumn,
]{revtex4-1}

\usepackage{graphicx}% Include figure files
\usepackage{dcolumn}% Align table columns on decimal point
\usepackage{subfigure}
\usepackage{bm}% bold math
\usepackage[utf8]{inputenc}
\usepackage{mathrsfs}
\usepackage[T1]{fontenc}
\usepackage[mathlines]{lineno}% Enable numbering of text and display math
\usepackage{mathrsfs}
\usepackage{multirow}
\usepackage{amsmath}
\usepackage{amssymb}
\usepackage{graphicx}
\usepackage{subfigure}
\usepackage{esint}
\usepackage{gensymb}
\usepackage{float}
\usepackage{lineno}
\usepackage{setspace}

\providecommand{\tabularnewline}{\\}
\makeatother
\usepackage{babel}

\begin{document}
%\linenumbers
\title{The linear and nonlinear inverse Compton scattering between microwaves and electrons in a resonant cavity}
\begin{spacing}{2.0}
\author{Meiyu Si}
\affiliation{Institute of High Energy Physics, Chinese Academy of Sciences, Beijing 100049, China}
\affiliation{University of Chinese Academy of Sciences, Beijing 100049, China}
\author{Shanhong Chen}
\affiliation{Institute of High Energy Physics, Chinese Academy of Sciences, Beijing 100049, China}
\affiliation{University of Chinese Academy of Sciences, Beijing 100049, China}
\author{Yongsheng Huang}
\email{ huangysh59@mail.sysu.edu.cn}
\affiliation{Institute of High Energy Physics, Chinese Academy of Sciences, Beijing 100049, China}
\affiliation{School of Science, Shenzhen Campus of Sun Yat-sen University, Shenzhen 518107, P.R. China}
\author{Manqi Ruan}
\affiliation{Institute of High Energy Physics, Chinese Academy of Sciences, Beijing 100049, China}
\author{Guangyi Tang}
\affiliation{Institute of High Energy Physics, Chinese Academy of Sciences, Beijing 100049, China}
\author{Xiaofei Lan}
\affiliation{Physics and Space Science College, China West Normal University, Nanchong 637009, China}
\author{Yuan Chen}
\affiliation{National Synchrotron Radiation Laboratory, University of Science and Technology of China, Hefei 230029, China}
\author{Xinchou Lou}
\affiliation{Institute of High Energy Physics, Chinese Academy of Sciences, Beijing 100049, China}
\affiliation{State Key Laboratory of Particle Detection and Electronics, Institute of High Energy Physics, CAS, Beijing 100049, China}

\begin{abstract}
The new scheme of the energy measurement of the extremely high energy electron beam with the inverse Compton scattering between electrons and microwave photons requires the precise calculation of the interaction cross-section of electrons and microwave photons in a resonant cavity. In the local space of the cavity, the electromagnetic field is expressed by Bessel functions. Although Bessel functions can form a complete set of orthogonal basis, it is difficult to quantify them directly as fundamental wave functions. Fortunately, with the Fourier expansion of Bessel functions, the local electromagnetic field can be considered as the superposition of a series of plane waves. Therefore, with corresponding corrections of the cross-section formula of the classical Compton scattering, the cross-section of the linear or nonlinear microwave Compton scattering in the local space can be described accurately. As an important application of our results in astrophysics, corresponding ground verification devices can be designed to perform experimental verifications on the prediction of the Sunyaev-Zeldovich(SZ) effect of the cosmic microwave background radiation. Our results could also provide a new way to generate wave sources with strong practical value, such as the terahertz waves, the ultra-violet (EUV) waves, or the mid-infrared beams.
\end{abstract}

\maketitle
\section{Introduction\label{sec:Introduction}}
The laser Compton scattering has a wide range of applications, such as the generation of high energy gamma beams in nuclear physics\cite{SPring, ELINP, shanghai, Yue Ma, Zhen-Chi, Xue-Kun}, the energy calibration of the electron beam\cite{Muchnoi, Tang, IHEP, Zhang} and the measurement of the beam polarization \cite{ILC, HERA}. Microwave inverse Compton scattering has become an interesting and valuable research topic for extremely high energy electron beams in colliders or space. However, the microwave Compton scattering process calculation is quite different from that of the traditional laser Compton scattering. In a traveling wave tube or a resonant cavity, the electromagnetic field is expressed by Bessel functions, which cannot be directly equivalent to the particle wave function for quantization. The new scheme of the energy measurement of extremely high energy electron beams with the inverse Compton scattering between electrons and microwave photons requires the precise calculation of the interaction cross-section of electrons and microwave photons in a resonant cavity\cite{Si}. 

The traditional quantum field theory is studied by quantization in the form of plane waves. Although Bessel functions can form a complete set of orthogonal basis, it is difficult to quantify them directly as fundamental wave functions. However, the Bessel functions in a local space can be expanded by the Fourier series. With corresponding corrections of the cross-section formula of the classical Compton scattering, the cross-section of the linear or nonlinear microwave Compton scattering in the local space can be described accurately.  The linear inverse Compton scattering is decided by the first-order term with our analytical results. Higher-order nonlinear inverse Compton scattering can occur because higher-order virtual photons are scattered by high energy electrons into high energy real gamma photons. The higher the order, the smaller the differential cross-section of the inverse Compton scattering between electrons and microwave photons. Therefore, the nonlinear inverse Compton scattering is much weaker than the linear one.  
The inverse Compton scattering between microwave photons and electrons also plays an important role in astrophysics studies\cite{1993Search}. In 1972, Sunyaev and Zeldovich proposed the  Sunyaev-Zeldovich (SZ) effect to describe the inverse Compton scattering between the cosmic microwave background(CMB) photons and the electrons in a galaxy cluster\cite{Hu2002COSMIC}. For the SZ effect, the CMB photons propagate as the plane waves in the universe.  Unequivocal detection of the SZ effect has profound cosmological significance\cite{1984, 1978, 1978White, 1979Birkinsha, 1979Fabbri, 1979}. The cross-section of the Compton scattering between microwave photons and electrons in a local space can be calculated, it is quite worthwhile to build a ground verification device to study or verify the SZ effect. In addition to the energy calibration of high energy electrons and the ground verification device for the SZ effect, the inverse Compton scattering of microwaves and electrons in a cavity will become a potential competitor of electromagnetic radiations in various frequency bands. For example, a terahertz wave with a frequency of 2 $\mathrm{THz}$ can be generated with microwaves scattered by an electron beam with the energy of 5 $\mathrm{MeV}$. The linear Compton scattering between microwaves and a 1 $\mathrm{GeV}$ electron beam\cite{2015, 2016} can also be an extreme ultra-violet (EUV) wave\cite{1973Ultraviolet} generator. The mid-infrared\cite{1991Mid} beams with the wavelength of three micrometers can be produced by the inverse Compton scattering between microwaves and an electron beam with the energy of 50 $\mathrm{MeV}$. These mid-infrared beams have wide application prospects and play irreplaceably important roles in the fields of remote sensing detection, environmental monitoring, biomedicine, scientific research, and optoelectronic countermeasures\cite{2007Remote,2013Mid,2014Mid, 2017Toward,2008Mid}. 
 
\section{Theoretical model in a resonant cavity}
For laser Compton scattering or the SZ effect in astrophysics, the Compton scattering cross-section is calculated using the traditional theory. The cylindrical resonant cavity is bounded, the existence of boundary conditions forms a standing wave field. The standing wave is stable, there is no movement of energy toward wave propagation.  There are three main resonant modes of TM$_{mnp}$, including TM$_{010}$, TE$_{011}$ and ${\rm TE_{111}}$. In ${\rm TM_{010}}$, $m=0,n=1,p=0$, the wavenumber $K=K_{\rm {c}}$=$\frac{V_{01}}{R}$, where $R$ is the radius of the cavity. The resonance wavelength $\lambda$=2$\pi /K$=2.613$R$. Figure 1 shows the resonant cavity.  The holes with a radius of $1.5\mathrm{mm}$ are made on the sidewall of the cavity to let the electron beam passes through. The inverse Compton scattering occurs between electrons and microwave photons.
\begin{figure}[ht]
\centering\includegraphics[scale=0.4]{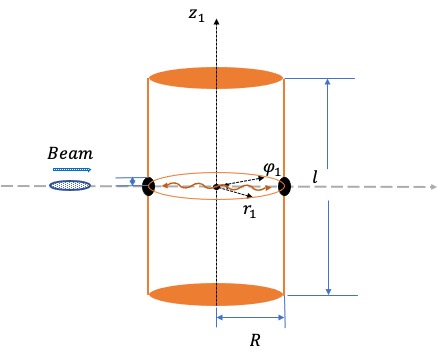}
\caption{The electron beam passes through the sidewall of a resonant cavity and undergoes inverse Compton scattering with microwave photons. In the cylindrical coordinate system ($r_{1}$,$\varphi_{1}$,$z_{1}$), $R$ is the radius of the cavity and $l$ is the length of the cavity. The holes with a radius of 1.5 $\mathrm{mm}$ are made on the sidewall of the cavity to let the electron beam pass through.}
\end{figure}
For the ${\rm TM_{010}}$ mode in a cylindrical coordinate system, the electromagnetic field is rotationally symmetric\cite{Microwave2}. The electromagnetic field can be written as follows: 
\begin{equation}
E_{z_{1}}=E_{m}J_{0}(K{\rm _ c}r_{1})e^{j\omega t},
\end{equation}
\begin{equation}
H_{\varphi_{1}}=jE_{m}\frac{1}{\eta}J_{1}( K{\rm _ c} r_{1})e^{j\omega t},
\end{equation}
\begin{equation}
E_{r_{1}}=E_{\varphi_{1}}=H_{r_{1}}=H_{z_{1}}=0,
\end{equation}
where $\eta=\sqrt {\frac {\mu_{0}}{\varepsilon_{0}}}$, the vacuum permeability $\mu_{0}=4\pi\times10^{-7} \mathrm{H/m}$,  the vacuum dielectric constant $\varepsilon_{0}=8.854188\times10^{-12} \mathrm{F/m}$. With Eq.(1), Eq.(2), and Eq.(3), the electric field only has the longitudinal field in $z_{1}$ direction, the magnetic field only has the transverse field in the $\varphi_{1}$ direction. In the Ref.\cite{Si}, the wavelength of microwave photons is 3.04 centimeters, the radius of the cavity is 11.65 $\mathrm{mm}$ and the field intensity of $E_{m}$ is $1\times 10^{7} \mathrm{V/m}$. The Poynting vector is the energy flow density vector in the electromagnetic field. The direction of the Poynting vector is radial, which also represents the motion direction of microwave photons.  Therefore, the inverse Compton scattering occurs between electrons and microwave photons as the electron beam passes through the resonant cavity from the sidewall. With Eq.(1) and Eq.(2), the electromagnetic field is expressed by Bessel functions, which are difficult to quantify as fundamental wave functions. However, the Bessel functions of a local space can be expanded by the Fourier series. The fields of Bessel functions are equivalent to the superposition of a series of plane waves.

\subsection{Fourier expansion of the microwave field in the cavity}
With Eq.(1), the electric field contains zero-order Bessel function, which is an even function. The $J_0 (x)$ can be expanded into a cosine series $\frac{a_{0}}{2}+\sum_{n=1}^{\infty } a_{n}\mathrm{cos} nx$. The expression is\cite{wuchongshi}
\begin{equation}
	J_0(x)=\frac{1}{\pi}\int_{0}^{\pi}\mathrm{cos}{(x \mathrm{sin} \theta)}d\theta.
\end{equation}
The Fourier series expansion of the $J_{0}(x)$ is 
\begin{equation}
	\begin{split}
	J_{0}(x)=&0.428931+0.608484\mathrm{cos}(1x)-0.051398\mathrm{cos}(2x)+0.021227\mathrm{cos}(3x)\\
	&-0.011661\mathrm{cos}(4x)+0.007383\mathrm{cos}(5x)-0.005098\mathrm{cos}(6x)+0.003733\mathrm{cos}(7x)-\cdots.
	\end{split}
\end{equation}
The detailed process is in the Appendix. The $J_{1}(x)$ can be expanded into a sine series $\sum_{n=1}^{\infty } b_{n}\mathrm{sin}  nx$. The integral expression is\cite{wuchongshi}
 \begin{equation}
	J_1(x)=\frac{1}{\pi}\int_{0}^{\pi}\mathrm{cos}{(x \mathrm{sin} \theta-\theta)}d\theta.
\end{equation}
The expansion of the $J_{1}(x)$ can be written as
\begin{equation}
	\begin{split}
	J_{1}(x)=&0.608484\mathrm{sin}(1x)-0.102796\mathrm{sin}(2x)+0.063683\mathrm{sin}(3x)\\
	&-0.046642\mathrm{sin}(4x)+0.036917\mathrm{sin}(5x)-0.030589\mathrm{sin}(6x)+0.026129\mathrm{sin}(7x)-\cdots.
	\end{split}
\end{equation}
Figure 2(a) and Figure 2(b) show the integral expression and the series expansion of $J_{0}(x)$ and $J_{1}(x)$ respectively.
\begin{figure}[ht]
  \centering
  \subfigure{\includegraphics[width=2.6in]{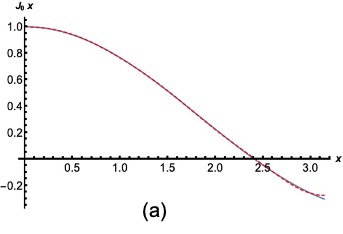}}
  \subfigure{\includegraphics[width=2.2in]{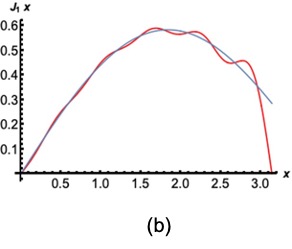}}
  \caption{(a) The zero-order Bessel function integral expression of Eq.(4) and the series expansion of Eq.(5) in the range of 0 to $\pi$. The red dotted line is the Fourier series expansion. The blue line is the integral expression. (b) The integral expression of $J_{1}(x)$ and the series expansion of Eq.(7). The red line is the Fourier series expansion. The blue line is the integral expression of $J_{1}(x)$.}
\end{figure}

Therefore, the electromagnetic field in a resonant cavity can be expanded into the superposition of a series of plane waves. The zero-order, first-order, and higher-order terms of the Bessel function series expansion are analyzed respectively. The zero-order term is  $J_{0}(K{\rm _ c}r_{1})^{(0)}=0.428931$, $J_{1}(K{\rm _ c}r_{1})^{(0)}\!=\!0$. According to the Eq.(1) and Eq.(2), the zero-order term of electromagnetic fields is $E_{z_{1}}^{(0)}\!=\!0.428931E_{m}e^{j\omega t}$, $H_{\varphi_{1}}^{(0)}=0$. The electric field is a constant. The first-order term of the Bessel function is $J_{0}(K{\rm _ c}r_{1})^{(1)}\!=\!0.608484 {\mathrm{cos}} (K{\rm _ c}r_{1}) $, $J_1(K{\rm _c}r_{1})^{(1)}\!=\!0.608484\mathrm{sin}(K{\rm _ c}r_{1})$. The exponential expression is
\begin{equation}
	J_{0}(K{\rm _ c}r_{1})^{(1)}=0.608484e^{-jK{\rm _ c}{r_{1}}},
\end{equation}
\begin{equation}
	J_{1}(K{\rm _ c}r_{1})^{(1)}=-0.608484e^{-j({\frac{\pi}{2}}+{K{\rm _ c}r_{1})}}.
\end{equation}
With Eq.(1), Eq.(2), Eq.(8) and Eq.(9),  the electromagnetic field of the first-order term is
\begin{equation}
	E_{z_{1}}^{(1)}={\rm F_{ (TM_{010})}^{(1)}}E_{m}e^{j(\omega t-K{\rm _ c}r_{1})},
\end{equation}
\begin{equation}
	H_{\varphi_{1}}^{(1)}=-{\rm F_{ (TM_{010})}^{(1)}}E_{m}\frac{1}{\eta}e^{j(\omega t-K{\rm _ c}r_{1})}.
\end{equation}
The coefficient of ${\rm F_{ (TM_{010})}^{(1)}}$ is equal to 0.608484. For $\overrightarrow{E}=-\frac{\partial{\overrightarrow{A}}}{\partial{t}}$, the vector $\overrightarrow{A}$ is given by

\begin{equation}
	A_{z_{1}}^{(1)}=-{\rm F_{ (TM_{010}))}^{(1)}}E_{m}\frac{1}{\omega j}e^{j(\omega t-K{\rm _ c}{r_{1}})}.
\end{equation}
$B_{\varphi_{1}}=-\bigtriangledown_{r_{1}}\times A_{z_{1}}$, $H_{\varphi_{1}}^{\prime}=B_{\varphi_{1}}/\mu_{0}=\frac{1}{\mu_{0}}(-\bigtriangledown_{r_{1}}\times A_{z_{1}})$. The dispersion relation of the first-order term is $\omega=K{\rm _ c}c$. The $H_{\varphi_{1}}^{\prime}$ is
\begin{equation}
	H_{\varphi_{1}}^{\prime}=-\frac{1}{\mu_{0}}\left(-{\rm F_{ (TM_{010})}^{(1)}}E_{m}\frac{1}{\omega j}e^{j(\omega t-K{\rm _ c}{r_{1}})}\right)\cdot {(-jK{\rm _ c})}=-\frac{1}{\mu_{0}c}{\rm F_{ (TM_{010})}^{(1)}}E_{m}e^{j(\omega t-K{\rm _ c}{r_{1}})}.
\end{equation}
For $\mu_{0}\varepsilon_{0}=\frac{1}{c^{2}}$, the relationship of $\frac{1}{\mu_{0}c}=\sqrt{\frac{\varepsilon_{0}}{\mu_{0}}}=\frac{1}{\eta}$ can be obtained. Simplify Eq.(13) to be
\begin{equation}
	H_{\varphi_{1}}^{\prime}=-{\rm F_{ (TM_{010})}^{(1)}}E_{m}\frac{1}{\eta}e^{j(\omega t-K{\rm _ c}{r_{1}})}=H_{\varphi_{1}}^{(1)},
\end{equation}
where the Eq.(14) is equal to Eq.(11). The expansion of the first-order electromagnetic field is satisfying Maxwell's equations. For the $E_{z_{1}}^{(1)}$ and $H_{\varphi_{1}}^{(1)}$, the frequency and wavelength are $\omega,\frac{2\pi}{K{\rm _ c}}$ respectively. The energy and momentum are $\hbar\omega$ and $\hbar $$\overrightarrow {K{\rm _ c}}$. The expression of $\omega^{2}=c^{2}K{\rm _ c}^{2}$ is the on-shell condition and the vacuum dispersion relationship. The linear inverse Compton scattering can occur between electrons and microwave photons.
For higher-order terms, the expansion of the Bessel functions are
\begin{equation}
 \begin{split}
	J_{0}(K{\rm _ c}r_{1})^{(2)+(3)+(4)+(5)+\cdots}=-0.051398\mathrm{cos}(2K{\rm _ c}r_{1})+0.021228\mathrm{cos}(3K{\rm _ c}r_{1})\\-0.011661\mathrm{cos}(4K{\rm _ c}r_{1})+0.007383\mathrm{cos}(5K{\rm _ c}r_{1})+\cdots
\end{split}
\end{equation}
and
\begin{equation}
\begin{split}
	J_{1}(K{\rm _ c}r_{1})^{(2)+(3)+(4)+(5)+\cdots}=-0.102796\mathrm{sin}(2K{\rm _ c}r_{1})+0.063683\mathrm{sin}(3K{\rm _ c}r_{1})\\-0.046642\mathrm{sin}(4K{\rm _ c}r_{1})+0.036917\mathrm{sin}(5K{\rm _ c}r_{1})+\cdots.
\end{split}
\end{equation}
The wave number is $nK{\rm _ c}(n>1)$, the resonant angular frequency $\omega$ remains unchanged. Since the energy of the photons is $\hbar\omega$ and the momentum is $\hbar n $$\overrightarrow{K{\rm _ c}}$, $\omega^{2}-(c\cdot nK{\rm _ c})^{2}<0$, the photons are virtual photons. The nonlinear inverse Compton scattering occurs between the virtual microwave photons and electrons. The detailed proof is given in the Appendix.

\subsection{The cross-section of the linear inverse Compton scattering}
According to the analysis in Section 2.1, the linear inverse Compton scattering occurs in the first-order term of the expansion of the field in a resonant cavity, while the nonlinear inverse Compton scattering occurs in the higher-order terms corresponding to virtual microwave photons. The corresponding interaction cross-section are required to calculate the flux of the scattered photons. The relationship is $\frac{{\rm d}N_{\gamma}}{d\omega^{\prime} {\rm d}t}=L\frac{d\sigma}{d\omega^{\prime}}$, where the $\frac{d\sigma}{d\omega^{\prime}}$ is the interaction differential cross-section, $L$ is the luminosity representing the ability to generate events. The cross-section of the inverse Compton scattering in a free space is\cite{2018FCC} 
 \begin{equation}
d\sigma_{0}=\frac{r_{e}^{2}}{\kappa^{2}(1+u)^{3}}\{\kappa[1+(1+u)^{2}]-4\frac{u}{\kappa}(1+u)(\kappa-u)\}dud\varphi,
\end{equation}
where $\kappa(\alpha)=4\frac{\omega_{0}\varepsilon_{0}}{(mc^{2})^{2}}\mathrm{sin}^{2}(\frac{\alpha}{2})$, the dimensionless constant $u=\frac{\omega^{\prime}}{\varepsilon_{0}-\omega^{\prime}}$. The $\varepsilon_{0}$ and $\omega_{0}$ is the initial energy of electron and the microwave photon respectively. The $\varepsilon $ and $\omega^{\prime}$ is the energy of the scattered electron and the scattered photon respectively. $\alpha$ is the collision angle between the initial electrons and the initial microwave photons. According to the first-order coefficients in electric and magnetic fields, the inverse Compton scattering cross-section in the cavity has a coefficient $|{\rm F_{ (TM_{010})}^{(1)}}|^{2}$. Therefore, the differential scattering cross-section in a resonant cavity is
\begin{equation}
d\sigma_{0}=|{\rm F_{ (TM_{010})}^{(1)}}|^{2}\cdot\frac{r_{e}^{2}}{\kappa^{2}(1+u)^{3}}\{\kappa[1+(1+u)^{2}]-4\frac{u}{\kappa}(1+u)(\kappa-u)\}dud\varphi,
\end{equation}
for $0<\varphi<2\pi$, the relationship between the differential cross-section and the energy of the scattered photons $\omega^{\prime}$ can be written as
\begin{equation}
\frac{d\sigma_{0}}{d\omega^{\prime}}=|{\rm F_{ (TM_{010})}^{(1)}}|^{2}\cdot 2\pi\frac{r_{e}^{2}}{\kappa^{2}(1+u)^{3}}\frac{\varepsilon_{0}}{(\varepsilon_{0}-\omega^{\prime})^{2}}\{\kappa[1+(1+u)^{2}]-4\frac{u}{\kappa}(1+u)(\kappa-u)\}.
\end{equation}
The microwaves with a wavelength of 3.04 $\mathrm{cm}$ collide head-on with 120 $\mathrm{GeV}$ electrons on CEPC. According to the inverse Compton scattering process\cite{Si}, the maximum energy of scattered photons is 9 $\mathrm{MeV}$. With Eq.(19), the relationship between the Compton differential scattering cross-section and the energy of scattered photons can be obtained in Figure 3.

\begin{figure}[h]
\centering\includegraphics[scale=0.4]{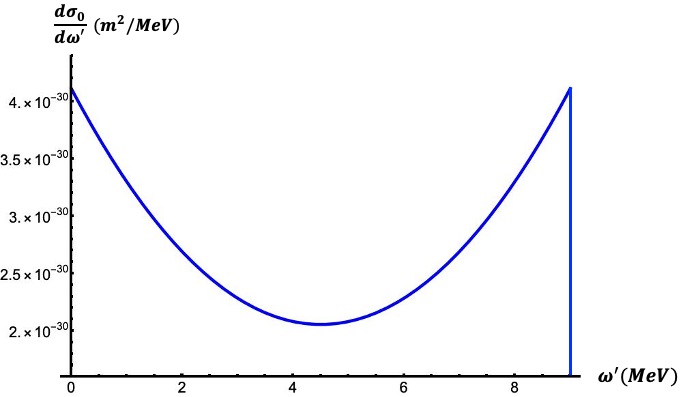}
\caption{For the microwaves with a wavelength of 3.04 $\mathrm{cm}$ collide head-on with 120 $\mathrm{GeV}$ electrons on CEPC, the maximum energy of scattered photons is 9 $\mathrm{MeV}$. The relationship between the Compton differential scattering cross-section and the energy of scattered photons in a local space is obtained by Eq.(19). The differential cross-section of electrons and microwave photons is about 0.04 $\mathrm{barn}$.\label{fig:The relationship between Compton differential scattering cross-section and the energy of scattered photons} }
\end{figure}

\subsection{The cross-section of the nonlinear inverse Compton scattering}
For the microwaves with a wavelength of 3.04 $\mathrm{cm}$, the initial microwave photons are virtual photons in higher-order terms shown by Eq.(15) and Eq.(16). With the conservation of energy and momentum in the inverse Compton scattering process, it can be found that the higher-order nonlinear inverse Compton scattering can occur because higher-order virtual photons are scattered by high energy electrons into high energy real gamma photons. For $n=2$, the maximum energy of the scattered photons is 14 $\mathrm{MeV}$; for $n=3$, the maximum energy of the scattered photons is 18 $\mathrm{MeV}$; for $n=4$, the maximum energy of the scattered photons is 23 $\mathrm{MeV}$; for $n=5$, the maximum energy of the scattered photons is 27 $\mathrm{MeV}$, and so on. According to the second-order electric and magnetic field coefficients, the nonlinear inverse Compton scattering cross-section in the cavity has a coefficient $\mid (-0.051398)\times(-0.102796)\mid$.
Therefore, the differential cross-section corresponding to the maximum energy of scattered photons are obtained and shown in Figure 4. The first point on the left in Figure 4 stands for the differential cross-section of the linear inverse Compton scattering. According to the expansion of the electromagnetic field in the cavity, the cross-section of the high-order terms is much smaller than those of the first-order term.  The higher the order, the smaller the differential cross-section between electrons and microwave photons. Therefore, the nonlinear inverse Compton scattering is much weaker than the linear one. 
\begin{figure}[ht]
\centering\includegraphics[scale=0.25]{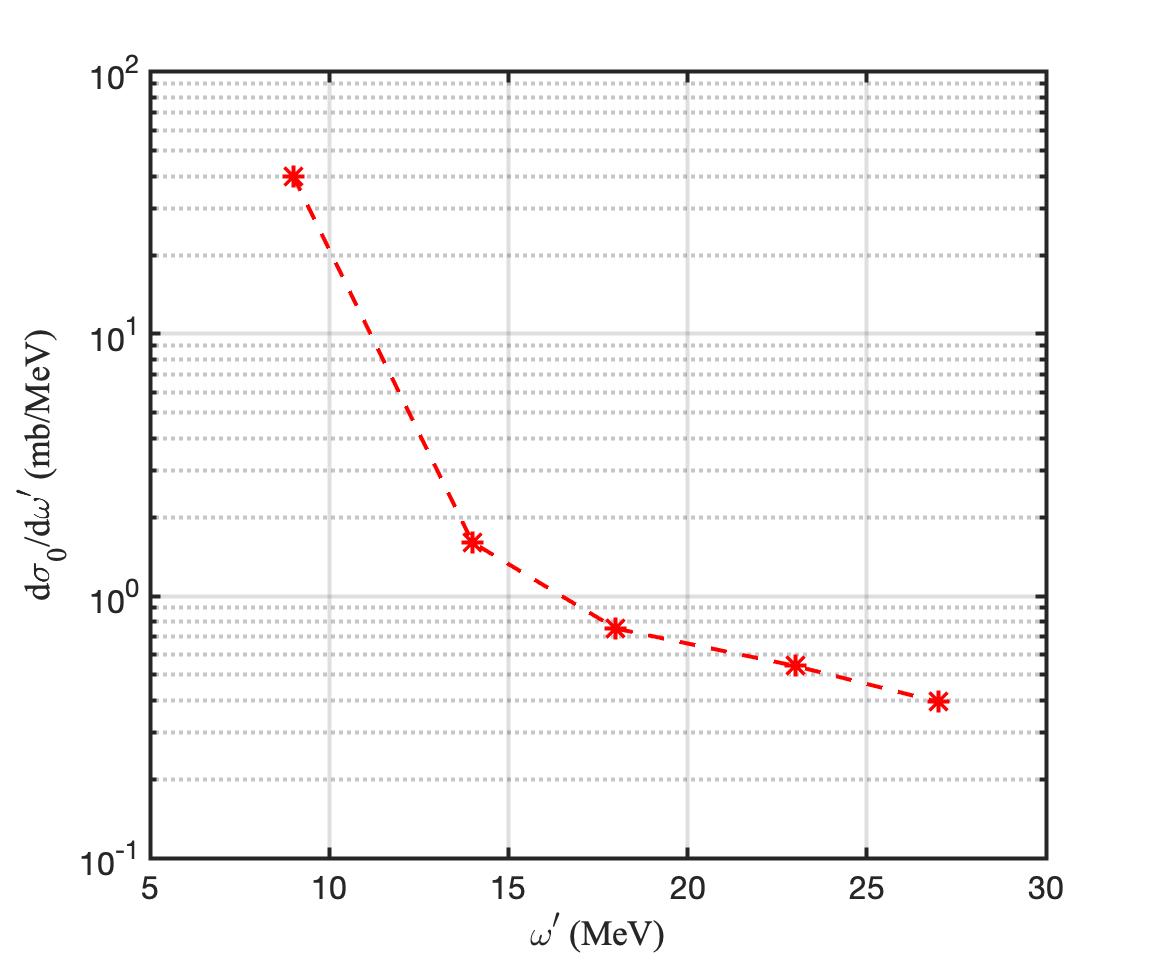}
\caption{
The differential cross-section of nonlinear Compton scattering between microwave photons with a wavelength of 3.04 $\mathrm{cm}$ and 120 $\mathrm{GeV}$ electrons in the head-on collisional mode.  According to the conservation of energy and momentum, the maximum energy of linear Compton scattered photons is 9 $\mathrm{MeV}$, and the high energy scattered photons correspond to the nonlinear inverse Compton scattering. The maximum energy of the nonlinear Compton scattering is $\omega^{\prime}$ = 14 $\mathrm{MeV}$, $\omega^{\prime}$ = 18 $\mathrm{MeV}$, $\omega^{\prime}$ = 23 $\mathrm{MeV}$, $\omega^{\prime}$ = 27 $\mathrm{MeV}$, corresponding to the nonlinear order 2, 3, 4, 5, respectively.}
\end{figure}

\subsection{An example: a new way for a far-infrared light source.}
The theoretical process of the inverse Compton scattering between electrons and microwave photons in a resonant cavity is calculated. The nonlinear inverse Compton scattering is much weaker than the linear one. The cross-section of the linear inverse Compton scattering in the cavity has a coefficient ${\rm |F_{(TM010)}^{(1)}|}^{2}$ compared with that in free space. To observe the inverse Compton scattering of electrons and microwave photons in a local space, we designed a simple experiment with the electron beam generated by a photocathode
electron gun. The beam parameters are shown in Table 1.
\begin{table}[h]
\centering
\begin{tabular}{cc}
\hline
  Parameters & Value \tabularnewline
\hline
 Beam energy  ($\mathrm{MeV}$) & 4-5\tabularnewline
 Bunch number  & 2  \tabularnewline
 Charge/bunch (nC) & >2  \tabularnewline
 Repeat frequency (Hz) & 50  \tabularnewline
\hline 
\end{tabular}
\caption{The parameters of the photocathode electron gun.}
\end{table} 
According to the inverse Compton scattering process in Ref.\cite{Si}, the maximum energy of the scattered photons is 0.015543 $\mathrm{eV}$ when the microwave photons with a wavelength of 3.04 $\mathrm{cm}$ collide head-on with 5 $\mathrm{MeV}$ electrons. The wavelength is 79.8 $\mathrm{\mu m}$. It belongs to the range of far-infrared light. Assuming the charge of a single bunch is 5 $\mathrm{nC}$, the number of electrons in a single bunch is about 3 $\times{10}^{10}$. The size of the bunch is taken as $\sigma_x$=$\sigma_y$=10 $\mathrm{\mu m}$, $\sigma_z$=3 $\mathrm{mm}$. With the calculation of the luminosity, the number of the maximum energy of scattered photons is 5165. The energy spectrum of scattered photons observed experimentally has a cutoff point at the maximum energy, called the Compton edge. It can be detected by a far-infrared spectrometer\cite{spectrometer}. Therefore, the maximum energy of the scattered photons is obtained by fitting the Compton edge. Assuming that a 0.05 Tesla magnet is used to deflect 5 $\mathrm{MeV}$ beam electrons, the critical energy of the synchrotron radiation photons can be calculated as follows:
\begin{equation}
\epsilon(\mathrm{keV})=0.665E^{2}(\mathrm{GeV})B(\mathrm{T})=8.31\times 10^{-7} \mathrm{keV}.
\end{equation}
Therefore, the synchrotron radiation photons generated by the bending magnet do not affect the detection of the scattered photons.

\section{Conclusion}
In a resonant cavity, the electromagnetic field is expressed by Bessel functions, which cannot be directly equivalent to the particle wave function
for quantization. Fortunately, the Bessel functions in a local space can be expanded by the Fourier series. By the Fourier expansion of Bessel functions, the electromagnetic field in the cavity can be expanded into the superposition of a series of plane waves. 
Therefore, based on the classical laser Compton scattering theory,  the cross-section of the microwave-electron inverse Compton scattering can be precisely predicted. According to the theoretical results, the first-order term decides the linear inverse Compton scattering of microwave photons and electrons. Higher-order nonlinear inverse Compton scattering can occur because high-order virtual photons are scattered by high energy electrons into high energy real gamma photons. The higher the order, the smaller the differential cross-section between electrons and microwave photons.  It can be applied to the simplification of the energy calibration of the electron beam with the energy of tens of or hundreds of $\mathrm{GeV}$\cite{Si} and the ground verification device for the SZ effect.

\section{Appendix}
The integral expression of the zero-order Bessel function is 
\begin{equation}
	J_0(x)=\frac{1}{\pi}\int_{0}^{\pi}\mathrm{cos}{(x \mathrm{sin} \theta)} d\theta.
\end{equation}
The $J_{0}(x)$ can be expanded into a cosine series $\frac{a_{0}}{2}+\sum_{n=1}^{\infty } a_{n}\mathrm{cos}  nx$. For $0<x<\pi$, the coefficients of $a_{0}$ and $a_{n}(n\ge 1)$ can be written as
\begin{equation}
	a_{0}=\frac{2}{\pi}\int_{0}^{\pi}J_{0}(x)dx=\frac{2}{\pi^{2}}\int_{0}^{\pi}\frac{\mathrm{sin}{(\pi \mathrm{sin}\theta)}}{\mathrm{sin}\theta}d\theta,
\end{equation}
\begin{equation}
	a_{n}=\frac{2}{\pi}\int_{0}^{\pi}J_{0}(x)\mathrm{cos}(nx)dx=\frac{(-1)^{n}}{\pi^{2}}\left[\int_{0}^{\pi}\frac{\mathrm{sin}{(\pi \mathrm{sin}\theta)}}{n+\mathrm{sin}\theta}d\theta-\int_{0}^{\pi}\frac{\mathrm{sin}{(\pi \mathrm{sin}\theta)}}{n-\mathrm{sin}\theta}d\theta\right].
\end{equation}
The $a_{0}$ and $a_{n}$ are calculated by numerical integration. Table 2 shows the numerical solution of Eq.(22) and Eq.(23).

\begin{table}[ht]
\centering
 \setlength{\tabcolsep}{0.5mm}{
\begin{tabular}{cccccccccc}
\hline
n & 0 & 1 & 2 & 3 & 4 & 5 & 6 & 7 &$\cdots$\tabularnewline
\hline
 $a_{n}$ & 0.857862  & 0.608484 &  $-0.051398$ & 0.021227 &  $-0.011661$ & 0.007383 & $-0.005098$ & 0.003733 &$\cdots$\tabularnewline
\hline 
\end{tabular}}
\caption{The numerical solution of the coefficients $a_{0}$ and $a_{n}(n\ge 1)$.}
\end{table}

The first-order Bessel function $J_{1}(x)$ in Eq.(2) is an odd function. The $J_{1}(x)$ can be expanded into a sine series $\sum_{n=1}^{\infty } b_{n}\mathrm{sin}  nx$. The integral expression of $J_1 (x)$ is\cite{wuchongshi}
 \begin{equation}
	J_1(x)=\frac{1}{\pi}\int_{0}^{\pi}\mathrm{cos}{(x \mathrm{sin} \theta-\theta)}d\theta.
\end{equation}
The expansion coefficient of $b_{n}(n\ge 1)$ is
\begin{equation}
	b_{n}=\frac{2}{\pi}\int_{0}^{\pi}J_{1}(x)\mathrm{sin}(nx)dx=\frac{1}{\pi^{2}}\left[\int_{0}^{\pi}\frac{\mathrm{cos}\theta -\mathrm{cos}(n\pi +\pi \mathrm{sin}\theta -\theta)}{n+\mathrm{sin}\theta}d\theta+\int_{0}^{\pi}\frac{\mathrm{cos}\theta -\mathrm{cos}(n\pi -\pi \mathrm{sin}\theta +
	\theta)}{n-\mathrm{sin}\theta}d\theta\right].
\end{equation}
Table 3 shows the numerical solution of $b_{n}$.
\begin{table}[h]
\centering
\begin{tabular}{ccccccccc}
\hline
n & 1 & 2 & 3 & 4 & 5 & 6 & 7 &$\cdots$\tabularnewline
\hline
 $b_{n}$ & 0.608484 &  $-0.102796$ & 0.063683 & $-0.046642$ & 0.036917 & $-0.030589$ & 0.026129 &$\cdots$\tabularnewline
\hline 
\end{tabular}
\caption{The numerical solution of the coefficient $b_{n}(n\ge 1)$.}
\end{table}

For the inverse Compton scattering of electrons and microwave photons, the conservation of the energy and momentum can be written as
\begin{equation}
\varepsilon_{0}+\omega_{0}=\varepsilon+\omega^{\prime},
\end{equation}
\begin{equation}
\vec p+\vec k_{0}=\vec p^{\prime}+\vec k,
\end{equation}
where ($\varepsilon_{0}$, $\vec p$), ($\omega_{0}$, $\vec k_{0}$) is the energy and momentum of the initial electrons and the initial photons respectively; ($\varepsilon $, $\vec p^{\prime}$), ($\omega^{\prime}$, $\vec k$) is the energy and momentum of the scattered electrons and the scattered photons respectively. A natural unit system with $\hbar=c=1$, the mass of the electrons is $m_{e}$. Therefore, the energy of the microwave photons is $\omega_{0}=K_{\rm c}$, the momentum is $\vec k_{0}=nK_{\rm c}$. For $n=1$, the microwave photons are real photons. For $n>1$, $\omega_{0}^{2}-(nK_{\rm c})^{2}<0$, the microwave photons are virtural photons.
Assuming that scattered photons are real photons, $(\omega^{\prime})^{2}=\vec k^{2}$. With Eq.(26), Eq.(27), the relationship can be written as follows:
\begin{equation}
\varepsilon_{0}-\sqrt{\varepsilon_{0}^{2}-m_{e}^{2}}+(1+n)K_{\rm c}=\varepsilon-\sqrt{\varepsilon^{2}-m_{e}^{2}}.
\end{equation}
For different values of $n$, the corresponding energy of scattered electrons and scattered photons can be obtained in Table 4. For $n=1$, the linear Compton scattering occurs of electrons and real microwave photons. For $n>1$, the nonlinear microwave-Compton scatterings between the electrons and virtual microwave photons happen and high energy scattered photons appear.
\begin{table}[h]
\centering
\begin{tabular}{ccccccc}
\hline
$n$ & 1 & 2 & 3 & 4 & 5  & 6\tabularnewline
\hline
$\varepsilon$ $(\mathrm{MeV})$ & 119991& 119986  & 119982 & 119977 &119973 & 119968 
\tabularnewline
$\omega^{\prime}$ $(\mathrm{MeV})$& 9 & 14  & 18 & 23 &27 & 32
\tabularnewline
\hline 
\end{tabular}
\caption{With the conservation of energy and momentum in the inverse Compton scattering, the corresponding energy of scattered electrons and scattered photons is obtained.}
\end{table}

\section{acknowledgment}
This work is supported in part by National Natural Science Foundation
of China (11655003); Innovation Project of IHEP (542017IHEPZZBS11820,
542018IHEPZZBS12427); the CAS Center for Excellence in Particle Physics
(CCEPP); IHEP Innovation Grant (Y4545170Y2); Chinese Academy of Science
Focused Science Grant (QYZDY-SSW-SLH002); Chinese Academy of Science
Special Grant for Large Scientific Projects (113111KYSB20170005);
National 1000 Talents Program of China; the National Key Research
and Development Program of China (No.2018YFA0404300).

\section{data availability}
The data that support the findings of this study are available from the corresponding author upon reasonable request.

\section{Disclosures}
The authors declare no conflicts of interest.

\section{CRediT authorship contribution statement}
Meiyu Si: Investigation, Validation, Formal analysis, Data curation, Writing – original draft. Yongsheng Huang: Conceptualization, Methodology, Formal analysis, Supervision, Project administration, Funding acquisition, Writing – review – editing. Shanhong Chen: Formal analysis. Manqi Ruan: Formal analysis.  Guangyi Tang: Formal analysis. Xiaofei Lan: Formal analysis. Yuan Chen: Formal analysis, Methodology. Xinchou Lou:  Formal analysis. 

%%%%%%%%%%%%%%%%%%%%%%% References %%%%%%%%%%%%%%%%%%%%%%%%%
%%%%%%%%%% If using BibTeX:
\bibliography{sample}

\begin{thebibliography}{}
 \vspace{3mm}
\bibitem{SPring} Ohkuma, Haruo, et al. "Production of MeV photons by the laser Compton scattering using a far infrared laser at SPring-8." Proceedings of EPAC. Vol. 2006.
\bibitem{ELINP} Suliman, Gabriel, et al. "Gamma beam industrial applications at ELI-NP." International Journal of Modern Physics: Conference Series. Vol. 44. World Scientific Publishing Company, 2016.
\bibitem{shanghai}  Luo, W., et al. "X-ray generation from slanting laser–Compton scattering for future energy-tunable Shanghai Laser Electron Gamma Source." Applied Physics B 101.4 (2010): 761-771.
\bibitem{Yue Ma} Ma, Yue, et al. "Region-of-interest micro-focus computed tomography based on an all-optical inverse Compton scattering source." Matter and Radiation at Extremes 5.6 (2020): 064401.
\bibitem{Zhen-Chi} Zhang, Zhen-Chi, et al. "Compact broadband high-resolution Compton spectroscopy for laser-driven high-flux gamma rays." Matter and Radiation at Extremes 6.1 (2021): 014401.
\bibitem{Xue-Kun} Xue, Kun, et al. "Generation of highly-polarized high energy brilliant $\gamma$ - rays via laser-plasma interaction." Matter and Radiation at Extremes 5.5 (2020): 054402.
\bibitem{Muchnoi} Muchnoi, N., Hans-Jürgen Schreiber, and Michele Viti. "ILC beam energy measurement by means of laser Compton backscattering." Nuclear Instruments and Methods in Physics Research Section A: Accelerators, Spectrometers, Detectors and Associated Equipment 607.2 (2009): 340-366.
\bibitem{Tang} Tang, Guangyi, et al. "The circular electron–positron collider beam energy measurement with Compton scattering and beam tracking method." Review of Scientific Instruments 91.3 (2020): 033109.
\bibitem{IHEP} An, G. P, et al. High energy and high brightness laser Compton backscattering gamma-ray source at IHEP. Matter and Radiation at Extremes, 3(4), 219-226.
\bibitem{Zhang} Zhang, Jian-Yong, et al. "Upgrade of beam energy measurement system at BEPC-II." Chinese Physics C 40.7 (2016): 076001.
\bibitem{ILC} E-166 Collaboration., and Achim W, Weidemann. "Polarized positrons at a future linear collider and the final focus test beam." International Journal of Modern Physics A 20.31 (2005): 7423-7432. 
\bibitem{HERA} Piotrzkowski, Krzysztof. "Measurement of the longitudinal polarization of the HERA electron beam using crystals and the ZEUS luminosity monitor." Nuclear Instruments and Methods in Physics Research Section B: Beam Interactions with Materials and Atoms 119.1-2 (1996): 253-258.
\bibitem{Si} Si, Meiyu, et al. "High energy beam energy measurement with microwave-electron Compton backscattering." Nuclear Instruments and Methods in Physics Research Section A: Accelerators, Spectrometers, Detectors and Associated Equipment (2021): 166216.
\bibitem{Microwave} K. Y. Zhao, "Microwave principle and technology.", Higher Education Press(2006).
\bibitem{1993Search} Boughn, S. P.. "Search for correlated large-scale structure in the cosmic x ray and cosmic microwave backgrounds." (1993).
\bibitem{Hu2002COSMIC} Wayne, Dodelson, and Scott. "Cosmic microwave background anisotropies.", Annual Review of Astronomy and Astrophysics (2002).
\bibitem{1984} M. rknshaw, S.F. u, and H. Hardck†. "The Sunyaev–Zeldovich effect towards three clusters of galaxies." Nature (1984).
\bibitem{1978} Gunn, J. E. ,  M. J. Rees , and  M. S. Longair . Observational cosmology. Cambridge University Press, 1978.
\bibitem{1978White} White, S. D. M., Silk, J. I. Astrophys. J. Lett. 226, L103-106 (1978).
\bibitem{1979Birkinsha} Birkinshaw, M. Mon. Not. R. astr. Soc. 187, 847-862 (1979).
\bibitem{1979Fabbri} Fabbri, R., Melchiorri, F. Mencaraglia, F., Natale, V. Astr. Astrophys. 74, L20-24 (1979).
\bibitem{1979} Cavaliere, A., Danese, L., De Zotti, G. Astr. Astrophys. 75, 322-325 (1979).
\bibitem{2015}Zhu, Xing-Long, et al. "Enhanced electron trapping and $\gamma$ ray emission by ultra-intense laser irradiating a near-critical-density plasma filled gold cone." New Journal of Physics 17.5 (2015): 053039.
\bibitem{2016}Zhu, Xing-Long, et al. "Dense GeV electron–positron pairs generated by lasers in near-critical-density plasmas." Nature communications 7.1 (2016): 1-8.
\bibitem{1973Ultraviolet} T Molde. "Ultraviolet Waves." South Australian Science Teachers Journal (1973):N/A.
\bibitem{1991Mid} Daly, J. G. . "Mid-infrared laser applications." Proceedings of SPIE - The International Society for Optical Engineering 1419(1991):94-99.
\bibitem{2007Remote} Fuchs, F., et al. "Remote sensing of explosives using mid-infrared quantum cascade lasers." Proceedings of SPIE - The International Society for Optical Engineering 6739.21(2007):1290 - 1292.
\bibitem{2013Mid} Nedeljkovi, M., et al. "Mid-infrared silicon photonic devices for sensing applications." 5th EOS Topical Meeting on Optical Microsystems International Society for Optics and Photonics, 2013.
\bibitem{2014Mid} Tittel, F. K., et al. "Mid-infrared Laser Based Gas Sensor Technologies for Environmental Monitoring, Medical Diagnostics, Industrial and Security Applications." Springer Netherlands (2014).
\bibitem{2017Toward} Hudson, D. D., et al. "Toward all-fiber supercontinuum spanning the mid-infrared." Optica 4.10(2017):1163.
\bibitem{2008Mid} Ebrahim-Zadeh, M., and  I. T. Sorokina. "Mid-Infrared Coherent Sources and Applications." NATO Science for Peace and Security Series B: Physics and Biophysics (2008):626.
\bibitem{Microwave2} H. G. Tang, C. Q. Tian, "Microwave Technology principle and Application analysis.", China Science and Technology Investment 000.036(2012):156-156.
\bibitem{wuchongshi} C. S. Wu, "Special topics on mathematical physics methods: mathematical equations and special functions.", Peking University Press, 2012.
\bibitem{2018FCC} N. Muchnoi, arXiv:1803.09595 [physics.ins-det] (2018).
\bibitem{spectrometer} Flasar F M, Kunde V G, Abbas M M, et al. Exploring the Saturn system in the thermal infrared: The composite infrared spectrometer[J]. Space Science Reviews, 2004, 115(1-4): 169-297.

\end{thebibliography}

%%%%%%%%%% If preparing manually:
 
\end{spacing}
\end{document}